\documentstyle[12pt]{article}
\begin {document}
\setcounter{page}{0}

\title{Optimization of the nonrelativistic potential model for $b\overline{b}$
quarkonia\footnote{Partly supported by the KBN grant 2P30207607}}

 \author{L. Motyka and K. Zalewski\thanks{Also at the Institute of Nuclear
 Physics, Krak\'ow, Poland}\\
 Insitute of Physics, Jagellonian University, Krak\'ow, Poland}

 \maketitle

\begin{abstract}
{A systematic study of a wide class of nonrelativistic models of
$b\overline{b}$ quarkonia is described. It is found that the potential $V(r) =
0.706380(\sqrt{r} - \frac{0.460442}{r}) + 8.81715$ (all in GeV) with the
$b$-quark mass $m_b = 4.80303$ GeV gives a satisfactory description of the
experimental data below the threshold for strong decays ($ \chi^2/DF = 6.5/7$).
Limitations and implications of this observation are discussed.}
\end{abstract}

\vspace{5cm}
\noindent TPJU - 7/95\\
March 1995
\newpage

Dozens of nonrelativistic and relativistically improved models of
$b\overline{b}$ quarkonia have been published. Many references can be found in
the recent review \cite{BES}. A fully relativistic treatment is beyond our
reach, even for the much simpler $e^+e^-$ system, but the folklore is that good
results can be obtained with a nonrelativistic potential, if the hamiltonian is
interpreted as an effective hamiltonian with the coefficients renormalized by
the relativistic corrections. Explicit relativistic corrections are necessary
to describe some purely relativistic effects, like the fine splitting. If,
however, one limits the discussion to centres of gravity of the multiplets,
than at the phenomenological level, as seen e.g. from \cite{BES}, the effective
nonrelativistic models are doing about as well as the relativistically improved
ones. In the literature a model is considered good, if it reproduces e.g. the
masses of the quarkonia within a few MeV. The experimental errors on the masses
today \cite{PDG}, however, are of the order of 0.2 MeV. This rises the
interesting question: is it possible to fit the data quantitatively, i.e.
within the experimental errors and if so, what is the corresponding effective
potential. The answer seems useful for at least the following two reasons.
Quark - antiquark potentials derived from studies of quarkonia are used in a
variety of applications. Let us mention as examples applications to heavy light
systems \cite{ISG}, to $b\overline{c}$ mesons \cite{FLS}, and to
$t\overline{t}$ production \cite{JEZ}. It is obviously advisable to use as
good potentials as possible and to know what are their limitations. On the
other hand, if it is not possible to fit the data with a nonrelativistic model,
this may be an interesting hint on how to construct a relativistic theory.

The first difficulty is that there is no such thing as a standard
nonrelativistic quarkonium model. The Schr\"odinger equation is, of course,

\begin{equation}
- \frac{1}{m_b}\vec{\nabla}^2 \psi(\vec{r}) + V(r) \psi(\vec{r}) =
E \psi(\vec{r}),
\end{equation}
but the quark mass $m_b$ and the potential $V(r)$ vary from paper to paper.
Our strategy is to limit the discussion to potentials of the form

\begin{equation}
\label{power}
V(r) = -a r^{-\alpha} + b r^\beta + Ct,
\end{equation}
where $a,b,\alpha, \beta, Ct$ are nonnegative constants. At least ten
potentials of this general form, but with various values of the parameters,
have been proposed in the literature. Thus the famous Cornell potential
\cite{EIC} has $\alpha = \beta = 1$. The potential advocated by Lichtenberg and
collaborators \cite{LIC} has $\alpha = \beta = 0.75$. The potential used by
Song and Lin in ref. \cite{SOL} has $\alpha = \beta = 0.5$, while Song's
potential used in ref \cite{SON} has $\alpha = \beta = \frac{2}{3}$. The
logarithmic potential of Quigg and Rosner \cite{QUR} corresponds to $\alpha =
\beta \rightarrow 0$. Potentials with $\alpha \neq \beta$ have also been
popular. Thus Martin \cite{MAR} has suggested $\alpha = 0,\; \beta = 0.1$,
while Grant, Rosner and Rynes \cite{GRR} prefer $\alpha = 0.045,\; \beta = 0$.
Heikkil\"a, T\"ornquist and Ono \cite{HTO} tried $\alpha = 1,\: \beta =
\frac{2}{3}$. Some very successful potentials known from the literature are not
of this type. Examples are the Indiana potential \cite{FOG} and the Richardson
potential \cite{RIC}. Our first observation is that these potentials are very
similar to the potentials considered by us, if their free parameters are fitted
to the data as explained below. This is illustrated in Fig. 1, where the
Richardson potential and the Indiana potential are compared with two potentials
of type (\ref{power}).

Let us make three comments concerning this figure. A figure analogous to our
Fig. 1 has been published long ago by Buchm\"uller and Tye \cite{BUT}. The
agreement between the curves on their plot was not as good as in Fig. 1, in
spite of the fact that they adjusted the constants in the potentials so as to
make all the potentials coincide in a chosen reference point. The reason is
that we have been much more selective, than was possible at their time, in the
choice of the "good" potentials for comparison. Our result supports the
conjecture of Quigg and Rosner \cite{QUR1}, who concluded from their inverse
scattering analysis that the first few $L=0$ energy levels and the
corresponding leptonic widths determine to a large extent the potential in the
region relevant for the quarkonium calculations. In our fits we used more
observables, but very similar results can be obtained by using just the masses.
Thus, a version of our analysis is similar to that from ref. \cite{QUR1},
except that instead of the leptonic widths we are using the centres of gravity
of the $L = 1$ states. Note, however, that we have no proof that a completely
different potential would not fit the data as well. Finally, let us stress that
in the regions of very large and very small values of $r$, not show in the
figure, the potentials are completely different from each other, but this has
practically no effect on the calculations for the quarkonia.

It is convenient to rewrite equation (\ref{power}) in the reduced form

\begin{equation}
-\vec{\nabla}^2 \phi(\vec{\rho}) + {\cal V}(\rho) \phi(\vec{\rho}) = {\cal E}
\phi(\vec{\rho}),
\end{equation}
where

\begin{eqnarray}
\vec{\rho} & = & \lambda \vec{r},\\
\lambda & = & \left(\frac{b}{a}\right)^{\frac{1}{\alpha + \beta}},\\
{\cal V}(\rho) & = & C( \rho^\beta - \rho^{-\alpha}),\\
{\cal E} & = & \frac{m_b}{\lambda^2}(E - Ct),\\
C & = & m_b a \lambda^{\alpha - 2}.
\end{eqnarray}
The idea is to concentrate on the observables, which depend on the parameter
$C$ only. We choose

\begin{eqnarray}
b_1 & = & \frac{M(2S) - M(1S)}{M(3S) - M(1S)}  =  0.6290 \pm 0.0005,\\
b_2 & = & \frac{M(3S) - M(2P)}{M(2S) - M(1P)}  =  0.774 \pm 0.006,\\
b_3 & = & \frac{M(2S) - M(1P)}{M(2S) - M(1S)}  =  0.219 \pm 0.001,\\
b_4 & = & \frac{|\psi_{2S}(\vec{0})|^2}{|\psi_{1S}(\vec{0})|^2}  =  0.492 \pm
0.111,\\
b_5 & = & \frac{|\psi_{3S}(\vec{0})|^2}{|\psi_{1S}(\vec{0})|^2}  =  0.433 \pm
0.071,\\
b_6 & = & |\psi_{1S}(\vec{0})|^{\frac{2}{3}}\langle 1P| r | 2S \rangle  =
2.29 \pm 0.16, \\
b_7 & = & |\psi_{1S}(\vec{0})|^{\frac{2}{3}}\langle 2P| r | 3S \rangle  =
1.59 \pm 0.15, \\
b_8 & = &  \frac{\langle 1S| r | 2P \rangle}{\langle 2S| r | 2P \rangle}  =
0.110 \pm 0.009.
\end{eqnarray}
Thus the $\chi^2$ distribution corresponds to seven degrees of freedom. All the
numerical values are calculated from the data given in the 1994 Particle Data
Group Tables \cite{PDG}.

This choice of observables requires some comments. Since our model is
nonrelativistic, we have replaced the masses of the $\chi$ states by the
centres of gravity of the multiplets. It would have been nice to be able to
include in the averagings also the masses of the $\eta_b$ and the ${}^1P_1$
states, but these masses are unknown. We have ignored the quarkonia with masses
above the threshold for strong decays. Their analysis would require coupled
channel calculations, which are much less well-defined. The extraction of
values of the wave functions at the origin $|\psi(0)|$ from the experimentally
measured leptonic decay widths is based on the Van Royen - Weisskopf formula
with a first order radiative correction (cf. e.g. \cite{BES}). Since the first
order correction is about 30\%, the second order correction is probably
significant. Unfortunately it is not known. Moreover, the formula itself leads
in certain cases to paradoxes (cf. \cite{NAZ} and references given there). It
is believed that the resulting uncertainties largely cancel in the ratios $b_4$
and $b_5$. In the observables $b_6$ and $b_7$, however, they introduce a
systematic error of perhaps some 7\%. This has not been included in our quoted
errors. Thus at this point we underestimate the confidence levels. Finally, the
relation between the dipole matrix elements occurring in the observables
$b_6,b_7,b_8$ and the measured dipole transitions is for quarkonia less close
than for atoms (cf. e.g. \cite{MOR}, \cite{MCB}). For our best fit, and for the
best fits to some other potentials, we have recalculated the full matrix
elements without the multipole expansion. This yields the correction factors,
which for transitions between $S$ and $P$ states are

\begin{equation}
C_{sp} = \frac{\sqrt{|<p|j_0(\frac{k r}{2})
\frac{d}{d r} |s>|^2 + 2 |<p|j_2(\frac{k
r}{2})\frac{d}{d r} |s>|^2}}{ |<p| \frac{d}{d r}
|s>|},
\end{equation}
where $s$ denotes the $S$-wave function, $p$ denotes the radial part of the
$P$-wave function, $j_l$ are the spherical Bessel functions and $k$ is the
length of the wave vector of the emitted photon. The corrections for the
transitions $2S \rightarrow 1P$ and $3S \rightarrow 2P$ related to $b_6$ and
$b_7$ are below $0.5$\%, i.e. negligible compared to the experimental
uncertainties. The ratio of the transition probabilities $(2P \rightarrow
1S)/(2P \rightarrow 2S)$ related to $b_8$ increases significantly, making our
fits worse, but not bad. The $\chi^2$ of the overall fit increases by about two
units.

As a first step we have calculated the observables $b_1,\ldots,b_8$ and the
corresponding values of $\chi^2$ for many of the existing models. Since our
comparison with experiment eliminates three of the four parameters and adjusts
the fourth to fit the data as well as possible, our agreement with experiment
is usually better than in the original papers. A representative selection of
the results is given in Table 1. We conclude that none of the models known to
us from the literature fits the data in the sense of the $\chi^2$ test.
Therefore, we have explored the quality of the fit in the region $0 \leq \alpha
\leq 1.2$, $0 \leq \beta \leq 1.1$ of the $\alpha,\beta$ plane. The resulting
map is shown in Fig 2. We have plotted the parameter $\log\frac{\chi^2}{7}$.
Thus the large variability of $\chi^2$ is clearly visible. We have also marked
the points corresponding to some of the models known from the literature. Good
fits correspond to the region $\chi^2 \leq 7$ i.e. to the region shaded on the
plot. All the models miss this region, though some come fairly close to it.

In order to get a good fit we choose in the shaded region a point (see Figure
2) with the simple coordinates $\alpha = 1,\; \beta = 0.5$. Minimizing $\chi^2$
with respect to the constant $C$ and then choosing the constants $\lambda$,
$m_b$, and $Ct$ so as to reproduce correctly the leptonic width of the
$\Upsilon(1S)$ state, i.e. $|\psi_{1S}(\vec{0})|^2$, the mass difference $M(3S)
- M(1S)$, and the mass $M(1S)$ and we find after substitutions

\begin{equation}
\label{vopt}
V(r) = 0.706380\left(\sqrt{r} - \frac{0.460442}{r}\right) + 8.81715,
\end{equation}
where $V(r)$ and $r^{-1}$ are in GeV. The constants are given with the
precision of six digits in order to assist the reader, who would like to check
our calculation. The corresponding quark mass

\begin{equation}
\label{masab}
m_b = 4.80303\mbox{ GeV}
\end{equation}
is quite reasonable. The corresponding predictions for the parameters
$b_1,\ldots,b_8$ are shown in the third line of the table. As expected the fit
is very good --- corresponding to a confidence level of 48\%. A more precise
treatment of the parameter $b_8$, as mentioned above, yield $\chi^2 = 8.8$
which corresponds to a confidence level of 27\%. The next best potential, the
Indiana potential \cite{FOG}, gets a similar correction, thus it does not
become competitive.

For $r \rightarrow 0$ our potential has the $r^{-1}$ dependence corresponding
to one gluon exchange. With present data, however, we have no evidence for the
additional factor $1/\log(\Lambda r)$, which according to QCD should be
introduced by the running of the coupling constant. The expected part of the
potential linear in $r$ is not seen. Probably the bottomonia are too small to
reach sufficiently far into the asymptotic region of linear confinement.
Perhaps a more flexible potential would exhibit the linear part.

Our conclusion is that it is possible to reproduce with a nonrelativistic
theory the observables $b_1\ldots,b_8$ within the experimental errors. The
corresponding potential (\ref{vopt}) and the corresponding estimate of the mass
of the $b$-quark (\ref{masab}) are very reasonable.

\newpage

\Large
\begin{center}
{\bf Table 1}\\ Comparison of the predictions of some models with experiment.
\end{center}
\normalsize

\noindent
\begin{tabular}{||l||c|c|c|c|c||c|c|c||c||}
\hline \hline
& $b_1$ & $b_2$   & $b_3$  & $b_4$ & $ b_5$ & $b_6$ & $b_7$
&$b_8$ & $\chi^2/DF$\\
\hline \hline
PDG data & 0.6290 & 0.7738 & 0.2187 & 0.49 & 0.43 & 2.31 & 1.59 & 0.110 &
\mbox{---}\\
\hline
error    & 0.0005 & 0.0057 & 0.0009 & 0.11 & 0.07 & 0.16 & 0.15 & 0.009 &
\mbox{---}\\
\hline \hline
This paper & 0.6293 & 0.7744 & 0.2191 & 0.49 & 0.36 & 2.26 & 1.37 & 0.124 &
 6.5/7 \\
\hline
Indiana & 0.6299 & 0.7829 & 0.2202 & 0.48 & 0.36 & 2.18 & 1.33 & 0.124 & 15.5/7
\\
\hline
Lichtenberg & 0.6283 & 0.7950 & 0.2172 & 0.48 & 0.36 & 2.19 & 1.34 & 0.126 &
 26.3/7 \\
\hline
Richardson & 0.6276 & 0.8106 & 0.2150 & 0.47 & 0.36 & 2.18 & 1.34 & 0.127 &
 74/7   \\
\hline
Song-Lin & 0.6382 & 0.7246 & 0.2375 & 0.49 & 0.35 & 2.04 & 1.21 & 0.112 & 850/7
\\
\hline
Cornell & 0.6128 & 0.8951 & 0.1946 & 0.47 & 0.37 & 2.42 & 1.55 & 0.142 & 2220/7
\\
\hline
Martin  & 0.6363 & 0.6891 & 0.2707 & 0.54 & 0.38 & 1.81 & 1.06 & 0.032 & 3720/7
\\
\hline \hline
\end{tabular}
\vspace{1cm}
\Large
\begin{center}
{\bf Figure captions}\\
\end{center}
\normalsize

{\bf Figure 1} Dependence of four typical potentials on the distance $r$. The
continuous curve correspond to potential (\ref{vopt}); the dotted curve
corresponds to the Indiana potential \cite{FOG} and to the potential
(\ref{power}) with $\alpha = \beta = 0.75$ \cite{LIC}, which coincide at the
scale of the figure; the dashed curve corresponds to the Richardson potential
\cite{RIC}. All the potentials have been scaled and shifted (see text) to fit
the data.

{\bf Figure 2} Map of the parameter $\log\frac{\chi^2}{7}$ in the $\alpha,
\beta$ plane. The region of very good fits ($\chi^2 < 7$) is shaded. The black
points correspond to: $\alpha = \beta = 1$ \cite{EIC}, $\alpha = \beta = 0.75$
\cite{LIC}, $\alpha = \beta = \frac{2}{3}$ \cite{SON}, $\alpha = 1,\; \beta =
\frac{2}{3}$ \cite{HTO}, $\alpha = \beta = 0.5$ \cite{SOL}, $\alpha = 0,\;
\beta = 0.1$ \cite{MAR}, $\alpha = 0.045,\; \beta = 0$ \cite{GRR}, $\alpha =
\beta \rightarrow 0$ \cite{QUR} and to $\alpha = 1,\;\beta = 0.5$ as proposed
in the present paper.

\end{document}